\documentclass[journal=jacsat,manuscript=article]{achemso}
\usepackage[version=3]{mhchem} 
\usepackage[T1]{fontenc} 
\usepackage[usenames,dvipsnames]{color}
\usepackage[hidelinks]{hyperref}

\author{Ashutosh Shukla}
\email{ashutosh.shukla@students.iiserpune.ac.in}
\author{Rahul Chand}

\author{Sneha Boby}

\author{G. V. Pavan Kumar}
\email{pavan@iiserpune.ac.in}
\affiliation[Unknown University]
{Department of Physics, Indian Institute of Science Education and Research Pune, Pune, Maharashtra, 411008, India}

\title{Synchronized motion of gold nanoparticles in an optothermal trap}


\begin{document}

\begin{abstract}
Optical tweezers have revolutionized particle manipulation at the micro- and nanoscale, playing a critical role in fields such as plasmonics, biophysics, and nanotechnology. While traditional optical trapping methods primarily rely on optical forces to manipulate and organize particles, recent studies suggest that optothermal traps in surfactant solutions can induce unconventional effects such as enhanced trapping stiffness and increased diffusion. Thus, there is a need for further exploration of this system to gain a deeper understanding of the forces involved. This work investigates the behaviour of gold nanoparticles confined in an optothermal trap around a heated anchor particle in a surfactant (CTAC) solution. We observe unexpected radial confinement and synchronized rotational diffusion of particles at micrometre-scale separations from the anchor particle. These dynamics differ from known optical binding and thermophoretic effects, suggesting unexplored forces facilitated by the surfactant environment. This study expands the understanding of optothermal trapping driven by anchor plasmonic particles. It introduces new possibilities for nanoparticle assembly, offering insights with potential applications in nanoscale fabrication and materials science.
\end{abstract}

\section{Introduction}
Optical tweezers have revolutionized our ability to control and study particles at the microscale and nanoscale, making them invaluable in biophysics, nanotechnology, and material science. \cite{ashkinObservationSinglebeamGradient1986,grierRevolutionOpticalManipulation2003,quidantOpticalManipulationPlasmonic2007, volpeRoadmapOpticalTweezers2023, spesyvtsevaTrappingMaterialWorld2016,dienerowitzOpticalManipulationNanoparticles2008a,roySelfAssemblyMesoscopicMaterials2013,royControlledTransportationMesoscopic2013,martinezColloidalHeatEngines2017,farreForceDetectionTechnique2010,deviTheoreticalInvestigationNonlinear2016} This technique uses focused laser beams to trap, move, and manipulate microscopic particles, such as biological cells, colloidal microparticles and nanoparticles. Optical trapping can adapt and integrate with various materials and geometries to improve control and efficiency. For instance, plasmonic trapping harnesses the resonance of conduction electrons in metallic nanoparticles to create stronger localized fields, enabling more efficient traps and enhanced precision at smaller scales.\cite{minFocusedPlasmonicTrapping2013,lehmuskeroLaserTrappingColloidal2015,liOpticalForcesInteracting2008,hongPlasmonicDielectricAntennas2024, crozierPlasmonicNanotweezersWhats2024,wangTrappingRotatingNanoparticles2011,ghoshAllOpticalDynamic2019,deviRevisitingNonlinearOptical2022,wang2011trapping}
Optical trapping near a surface, like a glass wall, can also significantly alter trap efficiency and dynamics.\cite{andrenSurfaceInteractionsGold2019,kimSurfactantsControlOptical2022,singhFluctuatingHydrodynamicsBrownian2017} These surface-mediated effects are critical for understanding particle-surface interactions, with applications like optical nanomotors\cite{lehmuskeroPlasmonicParticlesSet2014,shaoGoldNanorodRotary2015,figliozziDrivenOpticalMatter2017,brontecirizaOpticallyDrivenJanus2023,khanOpticallydrivenRedBlood2005,khanNanorotorsUsingAsymmetric2006} and optical printing.\cite{gargiuloAccuracyMechanisticDetails2017,liOpticalNanoprintingColloidal2019,ghoshDirectedSelfAssemblyDriven2021}
Optical tweezers can also induce optical binding between particles. This light-scattering mediated interaction can be used to manipulate multiple particles into stable self-assembled structures such as chains or arrays.\cite{qiOpticalBindingMetal2023,forbesOpticalBindingNanoparticles2020,yanPotentialEnergySurfaces2014,hanPhaseTransitionSelfStabilization2020, zhangDeterminingIntrinsicPotentials2024,huangPlasmonicDipoleQuadrupole2024} Conventional optical binding uses a large illumination area. Still, recent studies show indirect optical binding outside the focal spot of the laser. \cite{kudoSingleLargeAssembly2018,huangPrimevalOpticalEvolving2022,taoRotationalDynamicsIndirect2023}
\\
Optical tweezers can also utilize the localized heating of plasmonic particles and surfaces.  \cite{kolliparaOpticalManipulationHeats2023,chenHeatMediatedOpticalManipulation2022,patraLargescaleDynamicAssembly2016,jobyOpticallyassistedThermophoreticReversible2022a,juanPlasmonNanoopticalTweezers2011a,righiniSurfacePlasmonOptical2008,heldenDirectMeasurementThermophoretic2015} A thermal gradient at the microscale can help facilitate thermophoresis, thermo-osmosis, thermodiffusion, and thermoelectricity, among other auxiliary fields. These auxiliary fields provide additional control for particle manipulation at the nanoscale. Optothermal traps enable particle confinement at much lower laser powers than traditional optical tweezers, reducing the risk of damage to heat-sensitive materials and studying the interaction, dynamics, and assembly of different particles.\cite{linOptoThermoelectricNanotweezers2018, bregullaThermoOsmoticFlowThin2016,raniEvanescentOptothermoelectricTrapping2024,paulOptothermalEvolutionActive2022, franzlHydrodynamicManipulationNanoobjects2022,wangGrapheneBasedOptoThermoelectricTweezers2022, tiwariSingleMoleculeSurface2021,kotnalaOvercomingDiffusionLimitedTrapping2020,qianMicroparticleManipulationUsing2020, sharmaLargescaleOptothermalAssembly2020,singhUniversalHydrodynamicMechanisms2016a,haldarSelfassemblyMicroparticlesStable2012,panjaNonlinearDynamicsMicroparticle2024,caciagliControlledOptofluidicCrystallization2020a,munoz-martinezElectrophoreticDielectrophoreticNanoparticle2017}  In surfactant solutions, optothermal trapping has been shown to be even more efficient. This is also referred to as opto-thermoelectric (OTE) trapping. In OTE trapping, extra trapping strength comes from the charge separation of the surfactant's positive micellar and negative anionic parts in a thermal gradient. \cite{linOptoThermoelectricNanotweezers2018} This mechanism is discussed in detail in the Results section. Further, it was shown that surfactants control optical trapping near a glass wall by making positively charged layers that enhance the diffusion of trapped particles. \cite{kimSurfactantsControlOptical2022} By leveraging optothermal manipulation and surfactant-mediated interactions, it is possible to create novel nanoparticle assemblies at lower powers with varied applications.\cite{pengOptothermoelectricMicroswimmers2020,pengOptoThermophoreticManipulationConstruction2018,linReconfigurableOptothermoelectricPrinting2017, liOpticalNanoprintingColloidal2019}\\
Optothermal trapping has enabled diverse and efficient particle manipulation at the nanoscale. However, the mechanisms governing particle behaviour in surfactant-enhanced optothermal traps are not completely understood. While optothermal forces can attract and organize particles, introducing surfactants fundamentally changes trapping dynamics, creating unique interactions and behaviours not observed in standard optothermal or optical binding studies. In this study, we report an unconventional trapping behaviour of nanoparticles enabled by the presence of surfactant in the trapping environment.\\

We show the trapping of gold nanoparticles in an anchor particle-assisted trap in a surfactant solution. The observed dynamics are novel as particles get trapped outside the beam focus, and multiple trapped particles show synchronized diffusive motion with micrometre separation between them. We have previously demonstrated the ability of anchor particle-assisted traps to efficiently trap individual nanoparticles and nanoparticle assemblies.\cite{tiwariSingleMoleculeSurface2021,shuklaOptothermoelectricTrappingFluorescent2023a} In this study, we focus on the dynamics of particles in an anchor particle-assisted optothermal trap. We find that particles in solution are trapped at a radial position between 1 and 2 \(\mu m\) away from the drop-casted particle. Furthermore, trapped particles undergo rotational diffusion around the drop-casted particle. This rotational diffusion is highly synchronised for various particles, even with separations of around 2 to 4\(\mu m\) between the particles. As discussed in the Results section, neither temperature-induced thermophoresis nor optical binding satisfactorily explains the observed behaviour. We performed numerical simulations to investigate the influence of various conventional interparticle forces. We also studied how trapping behaviour varies with surfactant concentration, anchor particle size, and laser power.

\section{Methods}

\subsection{Materials}
Milli Q water was used to prepare the sample. Cetyltrimethylammonium chloride (CTAC) was obtained from Sigma-Aldrich in solution (25 wt\% in water). Gold nanoparticles with a 400 nm diameter, stabilised suspension in citrate buffer, were purchased from Sigma-Aldrich. Two kinds of suspensions were created with gold nanoparticles. One in water and the other in ethanol. 1 mL of nanoparticle suspension was centrifuged, and the supernatant was removed and redispersed in 2 mL of ethanol for drop-casting on the surface. Similarly, 1 mL of nanoparticle suspension was centrifuged and redispersed in 1 mL MilliQ water.
Two 50 µL solutions were prepared in micro-centrifuge tubes using MilliQ water to prepare nanoparticle suspension in surfactant solution. One with 5 µL of previously prepared aqueous gold NP suspension added to 45 µL water. Another tube had 10 µL of freshly prepared 50 mM CTAC concentration solution added to 40 µL of water. These 50 µL solutions were mixed to give the final solution, which was used for trapping.\\
Goldseal Cover Glass from Ted Pella, Inc. were used for experiments after cleaning with acetone. A double-sided adhesive spacer from Grace Bio-Laboratories (SecureSeal imaging spacer, 120 µm thickness) was used to build a microfluidic chamber. The spacer is pasted on the cover glass. 2 µL of the alcoholic gold suspension was drop-casted on the glass coverslip and allowed to dry by evaporation for 20 minutes. This yields gold NPs, which are firmly attached to the surface.  After 20 minutes, a 10 µL solution of the surfactant mixed with aqueous suspension is poured into the spacer, and the chamber is sealed with another cover glass from the top. 
\subsection{Optical Measurements}
The experiments were performed on a custom-built dual-channel microscope. The imaging was performed in brightfield mode, and trapping was performed in upright geometry. A 100$\times$, 0.95 NA objective lens (Olympus MPlanApo N) focused the laser from the top of the AuNP anchored on the cover glass. The same objective also illuminates the sample plane with white light from the top. Another 100$\times$, 1.49 NA objective lens (Olympus UApo N, oil immersion) collected all the transmitted light. The images were captured using a fast camera with a rate of 500 frames per second. The laser wavelength is filtered from the collected light before reaching the camera using a notch filter and a long pass filter. The setup was equipped with a three-dimensional piezoelectric translation stage for high-precision control of the sample.\\
A 532 nm linearly polarised Gaussian laser beam was used for the experiments. The power of the laser was controlled using a half-waveplate and a polarising beam splitter. This assembly provides the capability to control the power continuously and measure the power in the sample plane by calibration. The laser powers used are less than 15 mW in the sample plane. The beam was not expanded, and the back aperture of the objective was underfilled.
\subsection{Particle Tracking}
The images captured using the fast camera were tracked using the Trackmate plugin of the FiJi distribution of ImageJ\cite{tinevezTrackMateOpenExtensible2017,schindelinFijiOpenSourcePlatform2012}. The images were taken using brightfield illumination, and the gold nanoparticles were shown as dark objects on a bright background. These images were then inverted for tracking using the TrackMate plugin. The images extracted from the tracked processed video show particles as white objects on a black background.
\subsection{Electromagnetic Simulation}
We calculated the scattering forces between particles using a Python-based generalized multiparticle mie theory package, MiePy\cite{parker2020optical}. The forces were then integrated to get the potential profile (see Supporting Information S1). 
\subsection{Measurement and Calculation of Temperature and Optothermal Forces}
We characterize the temperature rise around the anchor particle used in our experiments numerically and experimentally, which match to a reasonable extent. The temperature increase around the anchor AuNP is estimated using a liquid crystal phase transition-based method \cite{franzlHydrodynamicManipulationNanoobjects2022,chand2023emergence} described in supporting information S2. Briefly, the anchor particle is heated in a liquid crystal environment, which leads to a nematic to isotropic phase transition above a temperature of 308 K. The phase boundary is measured and used to estimate the temperature rise on the particle's surface. On the other hand, we also estimated the temperature rise theoretically. The absorption cross-section of the nanoparticle was calculated using Mie theory, which was used to calculate the temperature rise of the nanoparticle.\cite{baffouThermoplasmonicsHeatingMetal2017a,hulst1981light} The temperature distribution around the nanoparticle was then calculated by solving the Heat-diffusion equation by convolving the heat source density with the Green's function. This is further described in detail in supporting information S3. Using the temperature rise of the nanoparticle, we conducted numerical simulations using a finite element method (FEM) solver to calculate the thermo-osmotic slip flow-induced force\cite{franzlHydrodynamicManipulationNanoobjects2022,chand2023emergence} (see Supporting Information S4). The OTE force is similarly computed using the temperature profile, as shown in Supporting Information S5.\cite{shuklaOptothermoelectricTrappingFluorescent2023a,linOptoThermoelectricNanotweezers2018}

\begin{figure}
\includegraphics[width=1\textwidth]{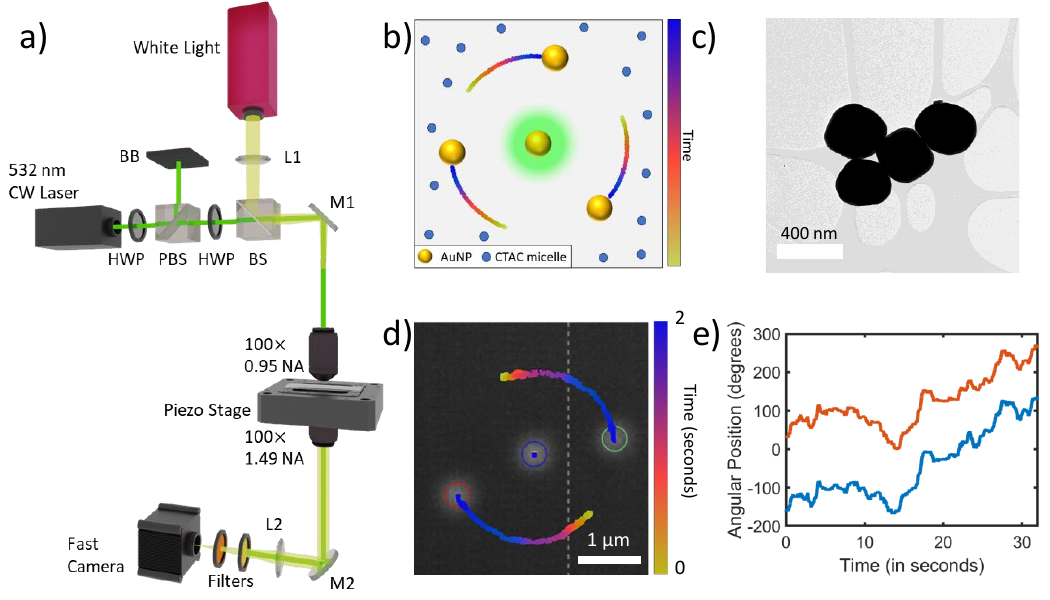}
\caption{Schematic figure. a) The optical schematic shows the dual-channel optical microscope setup, which consists of two objectives (100$\times$, 0.95 NA and 100$\times$, 1.4 NA) for excitation and signal collection, respectively. b) Schematic figure of the optothermal trap using 400 nm gold nanoparticles (AuNP) in the presence of surfactant (CTAC). c) Transmission Electron microscopy images of the used AuNPs. d) Experimental data corresponding to Supplementary Video 1 showing the trapping away from the centre and synchronised diffusion along the circumference. e) Angular trajectories of the trapped NPs showing synchronised rotational diffusion.}
\label{schema}
\end{figure}
\section{Results and Discussion}
We investigate the optothermal trapping of gold nanoparticles in a surfactant solution on a heated anchor particle-based optothermal trapping platform. To resolve the dynamics compared to our previous studies on trapping in surfactant solutions, we have performed the experiments on a dual channel setup as shown in the optical schematic in Figure \ref{schema}a. The setup is configured to collect light from the sample plane using a 100$\times$, 1.49 NA objective lens and captures images with a fast camera at 500 frames per second. The schematic Figure \ref{schema}b shows the experiment geometry. As described in the methods section, a 400 nm gold particle is drop-casted on a glass substrate and is firmly attached to the cover glass. We illuminate this particle, called the anchor particle, with the 532 nm laser using a high numerical aperture objective lens (100$\times$, 0.95 NA). The 400 nm AuNP has a significant absorption cross section at the used laser wavelength and leads to heating. The heated particle sets a temperature gradient on the glass substrate and the surrounding fluid, which can, in turn, set up a thermo-osmotic slip flow in the surrounding fluid. \cite{baffouThermoplasmonicsHeatingMetal2017a,franzlHydrodynamicManipulationNanoobjects2022,bregullaThermoOsmoticFlowThin2016,donnerPlasmonAssistedOptofluidics2011a} Gold nanoparticles diffusing in a surfactant solution follow the fluid flow and come near the anchor particle and get confined at a radial distance as shown schematically in Figure \ref{schema}b. Figure \ref{schema}c shows the transmission electron microscopy images of the 400 nm gold nanoparticles used for experiments.\\
Figure \ref{schema}d shows representative experimental observations corresponding to our schematic. The AuNP in the centre is drop-casted, and the AuNPs in the solution get trapped at around 1 \(\mu m\) distance far from the centre of the beam. These trapped particles perform rotational diffusion, which is visible in the 2-second trajectory plotted behind the particles. Throughout their motion, the particles are confined at a uniform radial separation from the anchor particle. Figure \ref{schema}e shows the angular coordinates around the particle for 30 seconds of rotational diffusion. It can be observed that the diffusion is synchronized as the angular trajectories move with each other. The initial angular separation between the particles is nearly 180 degrees, and the linear separation is of the order of 3 \(\mu m\).\\
We now describe the role of surfactant and anchor particles in optothermal trapping and discuss the observed synchronization in detail. We will also discuss the role of forces involved in the trapping process in the following subsections.
\subsection{Anchor Particle Driven Trapping}
\begin{figure}
\includegraphics[width= 1\textwidth]{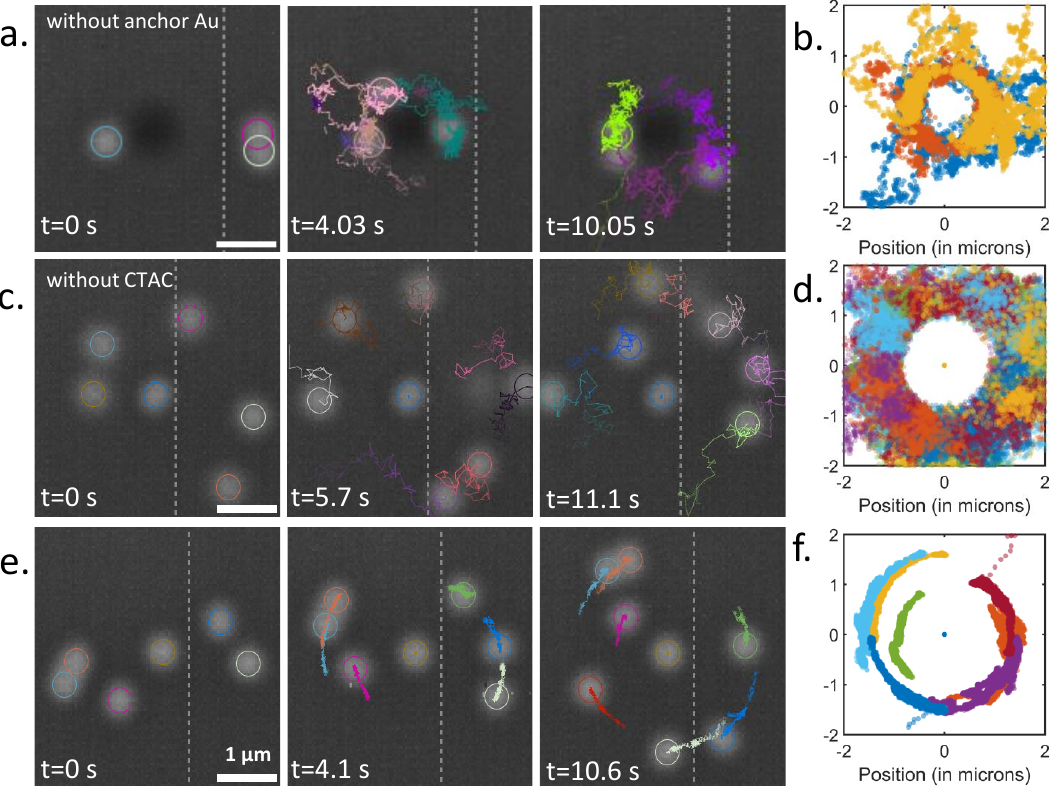}
\caption{Dynamics of gold nanoparticles in different traps (corresponding to the Supplementary video 2): a) without anchor gold particle, c) without surfactant (CTAC), and e) with both anchor particle and surfactant. The tracks of particles are shown for each condition in b), d), and f), respectively.}
\label{main}
\end{figure}
We note that the essential elements of the trap we observe are the anchor particle and the surfactant. We attempt to decompose the trapping process by individually considering the effects of these essential elements of the trap. We discuss these results systematically in Figure \ref{main}.  First, we discuss the trapping behaviour of particles in the surfactant solution without the anchor particle. The surfactant used in the experiments was Hexadecyltrimethylammonium chloride (CTAC), and the concentration used was always five millimolars (mM). The images are taken in transmission brightfield configuration as described in methods. In Figure \ref{main}a, the gold particles in the surfactant solution are confined around the laser beam. In this case, the beam is partially visible and creates a dark spot in the centre of the images in Figure \ref{main}a. The trajectories of the particles are random and Brownian, as shown in Figure \ref{main}b, with some attraction towards the beam. The particles hover near the beam but do not enter the high-intensity region of the beam owing to the transverse optical scattering force.\\
Secondly, Figure \ref{main}c shows the trap with the anchor particle without the surfactant. Here, the anchor particle heats up and sets up fluid flow that drives the particles in out-of-plane circuitous trajectories. The particles come radially inwards towards the anchor particle, diffuse out of the plane, and move radially outwards with the fluid flow. The particles again come in the sample plane at a larger radial distance and repeat the cycle. The trajectories of all particles correspond to this behaviour in Figure \ref{main}d.\\
Finally, in Figure \ref{main}e, we show the behaviour of trapped particles around the anchor particle in the presence of 5 mM CTAC. The AuNPs show remarkably different behaviour as different particles show confinement at various radii. The particles get confined radially and diffuse rotationally around the circumference. This is visible prominently in trajectories in Figure \ref{main}f. 
In all three cases, the particles are not trapped in the centre of the intensity gradient or the heat gradient following a Boltzmann distribution, as observed in previous studies.\cite{jonesOpticalTweezersPrinciples2015} Schmidt et. al. have previously shown that the position distribution of active nanoparticles in a critical solution held in an optical harmonic potential (785 nm laser beam) can transition from the Boltzmann distribution to a non-equilibrium state.\cite{schmidt2021non} We have not used a critical mixture, which is the main reason for their observation, and our particles are trapped more stably at a larger distance than their observations. Also, Tao et al. have shown, using simulations, that indirect optical illumination can hold 200 nm AuNPs outside the illumination region of a 1064 nm laser beam and lead to stable assemblies.\cite{taoRotationalDynamicsIndirect2023} Indirect optical binding is possible in their case because of the 1064 nm laser beam used with smaller particles. In our experimental study, the confinement behaviour of diffusing AuNPs at some fixed radii from the anchor AuNP is only observed in the presence of surfactant in the solution. The behaviour is absent if we remove either the surfactant or the anchor particle, as is shown in Figure \ref{main}a-d. The dominant force for offset trapping in all three cases discussed in Figure \ref{main} is the strong optical scattering force on the 400 nm AuNPs from the 532 nm laser beam. We further discuss this in supporting information S6. A comparison of the dynamics of particles around the anchor particle with and without surfactant is presented in Supporting Information S7. It shows that diffusion of the particles is suppressed in the surfactant-facilitated trap.\\
We have further investigated the role of the size of anchor particles in trapping by performing experiments with different anchor particles. A comparison of trapping using three different-sized anchor particles (150, 250, and 400 nm AuNPs) is shown in supplementary video 3. The mean trapping for all anchor particle radii lies between 1 and 1.5 \(\mu m\). There is a slight decreasing trend as the particle size is increased, which we also anticipate. Further details are analysed in supporting information S8. Similarly, the role of the surfactant concentration was studied by experimenting with the critical micellar concentration (cmc) of CTAC. Supplementary video 4 shows the comparison in trapping behaviour, and analysis is presented in supporting information S9. We observe stark contrast in mean trapping radii above and below the cmc of CTAC. The particles get trapped closer to the anchor particle below cmc, and the angular correlations are smaller as the concentration is reduced. \\
We have also studied the evolution of a trapped particle assembly as the laser power is changed. The mean radius of the trapped particles reduces steadily with the laser power. The result is discussed in depth in supporting information S10.

\subsection{Synchronised Motion of Trapped Particles}
\begin{figure}
\includegraphics[width= \textwidth]{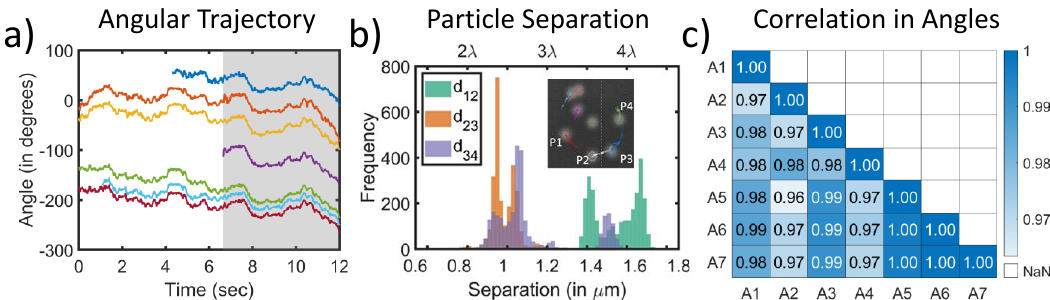}
\caption{Synchronised motion of trapped gold nanoparticles. a) Angular positions of trapped particles show synchronous displacements, as is also visible in Supplementary video 2. b) shows the distribution of linear separation d$_{12}$, d$_{23}$, and d$_{34}$ between the particles P1, P2, P3 and P4, respectively, as shown in the inset. The top four trajectories in a) correspond to P4, P3, P2, and P1. c) Correlation coefficients amongst all angular trajectories for the shaded region in a) are plotted in a heatmap. Angles are number A1 to A7 successively from the top in a).}
\label{angle}
\end{figure}
In the previous subsection, we described a gold nanoparticle's offset radial confinement and azimuthal diffusion around the anchor particle. When multiple particles are present in this trap, they show rotational diffusion around the circumference, which is synchronised, as shown in Figure \ref{angle}a. It shows the angular positions of the trapped particles in Figures 2e and f for around 12 seconds. The azimuthal diffusion is not completely coupled but varies slightly as the angular position tracks sometimes come closer. Figure \ref{angle}b shows the distribution of consecutive linear distances d$_{12}$, d$_{23}$, and d$_{34}$ between particles P1, P2, P3 and P4, as marked in the inset. The particle P1 is the last to join the assembly, and distances are measured after it has settled at a radial position. The histograms of the distances show that the linear distance between particles is not fixed but has peaks. This suggests a tendency to maintain a fixed separation from the surrounding particles. Still, there are also some deviations from that tendency, presumably due to thermal noise and local surface conditions. The consecutive interparticle distances range between 0.8 to 1.6 \(\mu m\) in this case.\\ 
We further calculated the Pearson correlation coefficients amongst the angular trajectories. We consider the part of trajectories where all particles have joined the trap. This time regime is highlighted in Figure \ref{angle}a. The correlation coefficients are visualised in a heatmap in Figure \ref{angle}c. The matrix elements (i, j) show the correlation coefficient between i-th and j-th angles, where angles in Figure \ref{angle}a are numbered successively from 1 to 7 from top to bottom. The Figure shows a very high correlation among the angles of the particles. This highly synchronised diffusion of particles with a micrometre order separation suggests some repulsive force of the micrometre range between particles. Other reports have shown such repulsion in different colloids due to the thermophoretic repulsion between heated particles\cite{paulOptothermalEvolutionActive2022} or the optical binding force\cite{hanPhaseTransitionSelfStabilization2020}. We suspect neither of these is responsible for the repulsion in this case. Thermophoresis is the tendency of particles to migrate in a temperature gradient.\cite{vigoloThermophoresisThermoelectricitySurfactant2010,rahmanThermodiffusionSoretEffect2014,burelbachUnifiedDescriptionColloidal2018} It is a surface phenomenon and the existence of a temperature gradient on the surface of the particle is essential to observe thermophoresis.\cite{andersonColloidTransportInterfacial1989} But owing to the high thermal conductivity of the gold nanoparticles, they have a uniform temperature across their surface. Consequently, they do not show thermophoretic repulsion. Moreover, thermophoretic repulsion creates assemblies where particles are equispaced around the circumference, as shown in a previous report\cite{paulOptothermalEvolutionActive2022}. In the current study, the particle separation is not uniform, as shown in Figure \ref{angle}b. We further present the detailed angular and linear separation analysis for a two-particle assembly (corresponding to Supplementary Video 5) in Supporting Information S11. The angular fluctuations are synchronised, but the absolute difference between angular position and linear separation can vary.\\
Meanwhile, for repulsion due to optical binding, a light scattering mediated interaction, the distance between the particles is integer multiples of the half-wavelength of the laser beam.\cite{forbesOpticalBindingNanoparticles2020} It is shown in Figure \ref{angle}b that the distance does not conform to that condition. Furthermore, our trapping beam, which has relatively low power, is focused on the anchor particles. The trapped particles are excluded from the high-intensity beam spot region. This contrasts with the broad beam used with higher powers for optical binding experiments, which is required to generate multiple scattering events of light between particles. 
\subsection{Numerical Estimation of Forces}
\begin{figure}
\includegraphics[width=\textwidth]{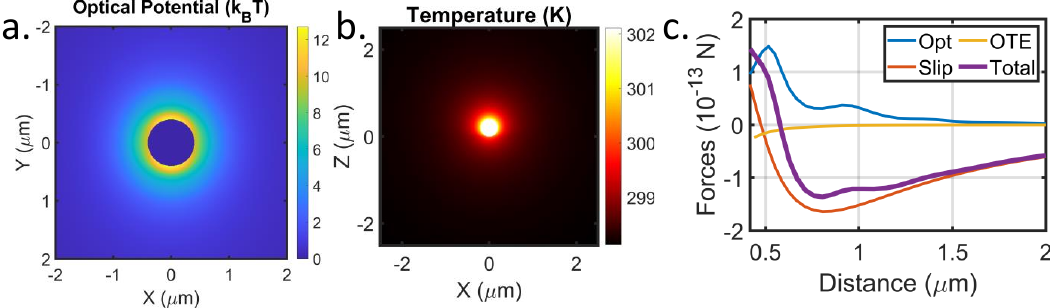}
\caption{Computation of Forces. a) Optical potential around the anchor AuNP calculated by integrating the optical scattering force. b) Temperature distribution around the heated anchor nanoparticle. c) The computed Optical, Opto-thermoelectric, and Thermo-osmotic slip induced drag forces on a particle near the anchor particle at various distances. The sum of all these forces along the x direction is also plotted to show the attractive behaviour at a large distance.}
\label{sims}
\end{figure}
In this section, we study numerically various forces present in our system which could give rise to the observed dynamics of particles. We focus on the observation of the confinement at a radial position. The major forces present in the system are the optical forces due to the laser beam, heating-induced opto-thermoelectric force, and thermo-osmotic slip flow-induced drag force. In the presence of the laser beam, the diffusing nanoparticles experience a force due to momentum exchange with the optical field. The resultant optical force is generally decomposed into scattering and gradient force, where the gradient force is responsible for the stable trapping of particles at the beam centre. In our study, the AuNPs used in the solution have a large scattering cross section and experience a large scattering force. Consequently, they are unable to access the high-intensity region of the beam. We show the optical potential profile around the anchor particle, shown in Figure \ref{sims}a. The potential is calculated at one mW of illumination power with a beam waist of around 1 \(\mu m\). The optical potential is positive at all places, indicating that the in-plane force on any diffusing particle is always repulsive, pointing away from the anchor particle. The optical potential is calculated using the generalized multiparticle Mie theory-based calculation described in the methods. The potential profile is not circularly symmetric as the excitation laser is x-polarised. This gives a large force along the y-direction as compared to the x-direction. \\
We also simulate the heating of the anchor particle by calculating the absorption coefficient of the gold nanoparticle using the Mie theory as shown in Figure \ref{sims}b. Our simulated temperature rise matches well with the experimental measurement (See Supporting Information S2 and S3) of the heating of anchor nanoparticles. We can then calculate the heating of the surrounding substrate and fluid, allowing us, in turn, to calculate the fluid flow around the particle. This fluid flow is classified into two parts. The first part is convection, caused by changes in the density of the heated fluid. The second part is thermo-osmotic slip flow, resulting from differences in the free energies of the fluid across a temperature gradient parallel to the substrate. We focus on thermo-osmotic slip flow as it is the dominant contribution. We modelled this thermo-osmotic slip flow using a commercial FEM-based solver using the expression from the literature $v_{||}=\chi \frac{\nabla _{||} T}{T}$, where $\chi$ is the thermo-osmotic slip coefficient, $T$ the temperature and $\nabla_{||}T$ is the temperature gradient parallel to the surface. We then calculate the hydrodynamic drag force on the particle from this fluid flow using the Stokes drag formula $F_{slip}=6\pi\eta R v_{||}$, where $\eta$ is the viscosity of the fluid, and $R$ is the radius of the particle.\\
We can also calculate the opto-thermoelectric (OTE) force using the temperature profile around the particle. OTE force is very prominent in optothermal trapping studies in the presence of the surfactant. The OTE force arises in systems with surfactants and local heating, as shown in the literature. The working principle of this force is the separation of charges of the cationic surfactant in the solution. The surfactant (CTAC) molecule dissociates into the chloride ion and the CTA\(^+\) cation in the solution. For concentrations exceeding the critical micellar concentration of the surfactant, the CTA\(^+\) ions self-assemble into micelles. The CTA\(^+\) ions are also adsorbed on all particle surfaces, rendering them positively charged. In the presence of the temperature distribution, the macro-cations (micelles) and the anions start moving down the temperature gradient owing to their intrinsic thermophoresis. However, the drift speed of migration of the macro-cations is larger than that of anions as the anions are smaller and more diffusive. This difference in the thermo-diffusivities results in a separation of charges in the steady state. This separation of charges creates an electric field which drives the positively charged particles towards the heat centre. See supporting information S5 for further details. This mechanism can provide a large force for stable trapping in many cases, but it is not significant in our case, as seen in Figure 4c.  The OTE force is directly proportional to the product of the temperature gradient and temperature. The temperature field decays as $\frac{1}{r}$ and so the temperature gradient decays as $\frac{1}{r^2}$ and thus the opto-thermoelectric force scales as $\frac{1}{r^3}$. Thus, the OTE force reduces in magnitude significantly compared to the other forces in the separation range of interest.\\
We combine all computed forces in Figure \ref{sims}c, which shows an equilibrium position near 0.6 \(\mu m\). Since the contribution of the OTE force is negligible, the force profile would be almost identical for the case without the surfactant, which has starkly different dynamics. Thus, we can infer that the surfactant facilitates some additional interactions which are not yet understood in the literature.\\
We would like to mention that due to the presence of the concentration gradient of surfactant micelles, there could also be a depletion interaction. However, this depletion interaction has been shown to be weaker than the OTE force in surfactant-assisted Optothermal traps. \cite{linOptoThermoelectricNanotweezers2018} Thus, we have not considered its contribution to our analysis.\\
To conclude, we would like to mention that the forces responsible for the synchronized motion are poorly understood. Hydrodynamic synchronization in micron-sized, dielectric, optically driven colloids has been shown in various contexts.\cite{kotar2010hydrodynamic,leyva2022hydrodynamic,miyamoto2019hydrodynamic} We believe such hydrodynamic synchronization could be the possible reason in our experiments as well. However, the AuNPs do not show any hydrodynamic correlation in the absence of the surfactant. Owing to a multitude of factors, such as the presence of laser scattering from the anchor particle, the surfactant micelle concentration gradient and the trapping close to the surface, where micelles are adsorbed, the problem here is more complex, and possibly a mixture of multiple effects to understand it. Since the trapped particles also experience laser illumination since the trapping radii are of the order of the beam waist,  there would also be finite heating of the trapped AuNPs. This heating would modify the thermo-osmotic flows, further complicating the analysis. Additionally, the system is highly charged due to the presence of CTA\(^+\) micelles and corresponding Chlorine anions. This high-charge environment could facilitate electrostatic repulsion-like force mediated by surfactant micelles.\\
In the following subsection, we discuss the experimental observation of rotation of the trapped assemblies and the dependence of rotation on experimental parameters.

\subsection{Rotation of Trapped Assembly}
\begin{figure}
\includegraphics[width=\textwidth]{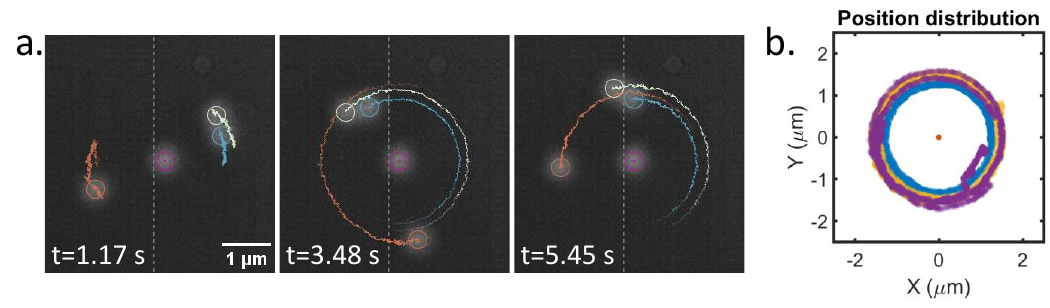}
\caption{Rotation of the trapped particle assembly. a) Time series of particles showing the emergence of spontaneous rotation in trapped particles. b) Trajectories of particles.}
\label{rotate}
\end{figure}
The nanoparticle assemblies that are formed tend to rotate around the anchor particle. This rotation can be seen in all supplementary videos, and the rotation direction is stochastic. In this section, we experimentally demonstrate that faster spontaneous rotation can emerge in the system. Figure \ref{rotate}a shows the time series of the particles corresponding to the supplementary video 6. The particles are trapped at radial positions, as described previously. As shown in Figure \ref{rotate}a, the three-particle system diffuses and gets into a particular configuration due to thermal noise. After achieving a particular configuration, the particle system starts rotating suddenly. The particle system rotates for two rotations in about 5 seconds. Then, the particles rearrange again due to fluctuations, and the rotation ceases. The position distribution in Figure \ref{rotate}b shows that the particles cover a circular trajectory.\\
Recently, Louis et al. have also shown trapping and correlated motion of 200 nm AuNPs using a 1064 nm laser beam.\cite{louis2024unconventional} In particular, they show correlated rotative oscillations for three trapped particles. In their case, stable optical trapping is possible because of the particle-laser combination they experimented with, viz. 200 nm diameter AuNP with a 1064 nm laser beam. This is similar to what Tao et al. used to show indirect optical binding.\cite{taoRotationalDynamicsIndirect2023} The particle separation in their experiments is of the order of a wavelength in the medium. In contrast, the particles in our experiments show synchronised motion at even separation of the order of 4 times the laser wavelength in the medium, as shown in Figure \ref{angle}. We also do not observe any stable conformation of the particles. All assemblies are formed stochastically based on the direction of the particle approach.\\
We observe this rotation using linearly polarised light, which carries no spin angular momentum (SAM). We further probed the effect of light with SAM on the assembly's rotation direction, as has been observed for individual particle plasmonic trapping in literature.\cite{wang2011trapping} We found that the angular momentum of the light does not influence the rotation direction of the assembly. A representative experiment is shown in supplementary video 7 and described in Figure \ref{circular}. The particles are trapped as described in the subsection `Anchor particle driven trapping' but using a circularly polarised laser beam. We use a quarter wave plate to make the incident light circularly polarised.\\
\begin{figure}
\includegraphics[width=0.7\textwidth]{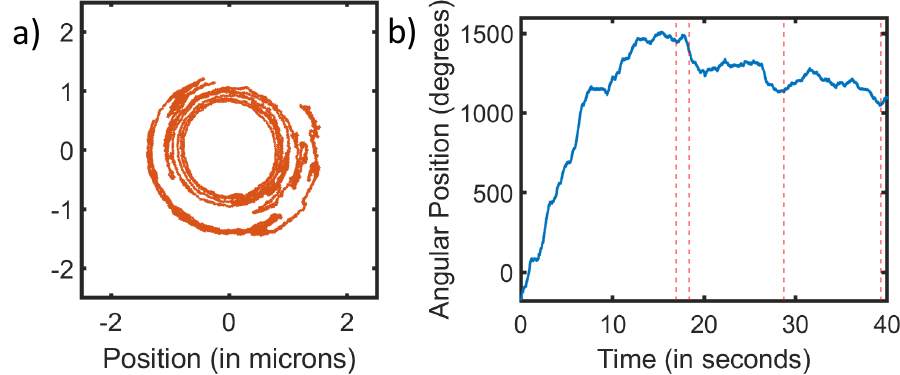}
\caption{Trapping with circularly polarised light. a) shows the trajectory of one trapped particle from the supplementary video 7 for 40 seconds. The angular position of the particle is shown in b). The red dotted lines mark the time points when a new particle joins the assembly. Using a fixed circular polarisation does not tend to induce rotation in a preferential direction.}
\label{circular}
\end{figure}
The trajectory of one trapped particle is shown in Figure \ref{circular}a instead of all particles in the assembly for clarity. The corresponding angular trajectory is shown in \ref{circular}b for 40 seconds. The assembly rotates in the counter-clockwise direction for more than four rotations in 15 seconds, then the rotation stagnates. More particles start joining the assembly, and the rotation direction slowly reverses. The multi-particle assembly performs roughly one clockwise rotation in the next 25 seconds. The red dotted lines in Figure \ref{circular}b are the time points when a new particle joins the assembly. Thus, the rotation direction of the assembly does not depend explicitly on the laser polarisation. Therefore, the observed rotation direction is stochastic and depends on the specific conformation of the assembly.

\section{Conclusion}
This study explores the dynamics of nanoparticles in anchor particle-based optothermal traps, focusing on the behaviour of 400 nm AuNPs in CTAC surfactant solution. We observed that nanoparticles are not trapped at the beam focus. Instead, we confined them at micrometer-scale distances from the heated anchor particle, where they exhibit synchronised rotational diffusion and spontaneous rotation. The presence of a repulsive interparticle force appears to regulate their motion, enabling synchronisation even at separations of several micrometres. The results highlight the critical role of CTAC in shaping nanoparticle interactions in an optothermal trap. Our results indicate that radial confinement and coordinated motion are absent in the absence of either the anchor particle or surfactant. This suggests that additional forces beyond known optical and thermophoretic mechanisms contribute to the observed trapping behaviour. Further studies are required to understand the precise mechanism of the synchronisation.

By studying these novel trapping dynamics, this work aims to expand our understanding of optothermal trapping in surfactant environments. Our experiments demonstrate that nanoparticle confinement and synchronised motion depend on key parameters such as surfactant concentration, anchor particle size, and laser power. We observed that reducing the surfactant concentration below the critical micellar concentration led to closer trapping near the anchor particle and weaker angular correlations. Varying the size of the anchor particle slightly altered the trapping radius, while increasing laser power steadily increased the mean trapping radius. We also observe the rotation of trapped particle assemblies. Experiments with circularly polarised light confirmed that the spontaneous rotation of particle assemblies is not driven by the optical spin angular momentum but instead emerges from the trapped assemblies’ specific conformations.\\
These results collectively highlight the complex interplay of optical, thermal, and surfactant-mediated forces, paving the way for new approaches and applications in materials science, nanofabrication, and optoelectronics, where controlled particle interaction and organisation are essential.
\begin{acknowledgement}
The authors thank Dr Adarsh Vasista, Dr Sunny Tiwari and Dr Diptabrata Paul for their valuable discussions on this project. We acknowledge using common experimental facilities in the Chemistry Department at IISER Pune. AS acknowledges the Ministry of Education, Government of India, for the Prime Minister Research Fellowship. This work was partially funded by AOARD (grant number FA2386-23-1-4054) and the Swarnajayanti fellowship grant (DST/SJF/PSA-02/2017-18) to G.V.P.K.
\end{acknowledgement}

\begin{suppinfo}
Supplementary information containing the following information and videos can be found at this \href{https://drive.google.com/drive/folders/1tN61cxpRY0Iqr9F3Sf1rcxoqLTuGcRu0?usp=sharing}{\color{blue} link}.\\ Computation of optical potential, Measurement of temperature using liquid crystal phase transition, Computation of temperature rise and distribution, Computation of thermo-osmotic slip flow and force, Computation of opto-thermoelectric force, Optical binding comparison, Comparison of particle distribution and dynamics with and without CTAC, Variation of anchor particle size, Variation of surfactant concentration, Laser power dependence on trapping behaviour, Two particles synchronised rotational diffusion (pdf)\\
Supplementary Video 1: Experiment video showing trapping and synchronisation of gold particles.
Supplementary Video 2: Comparison of trapping dynamics in various conditions. a) without anchor particle and with CTAC, b) without CTAC and with anchor particle, c) with anchor particle and five mM CTAC.
Supplementary Video 3: Experiment video comparing the trapping with variation of anchor particle size.
Supplementary Video 4: Experiment video comparing the trapping with variation of surfactant concentration.
Supplementary Video 5: Experiment video with simultaneous angular position graph showing synchronised motion.
Supplementary Video 6: Experiment video showing spontaneous rotation in trapped particle assembly.
Supplementary Video 7: Experiment video showing the effect of circularly polarised light in rotation of trapped particle assembly.
\end{suppinfo}

\bibliography{0OptothermalBinding}

\end{document}